# CRISP - WP11

## Moderator neutron cross-section data

### The neutron cross section of the hydrogen liquids:
#### substantial improvements and perspectives

ELEONORA GUARINI

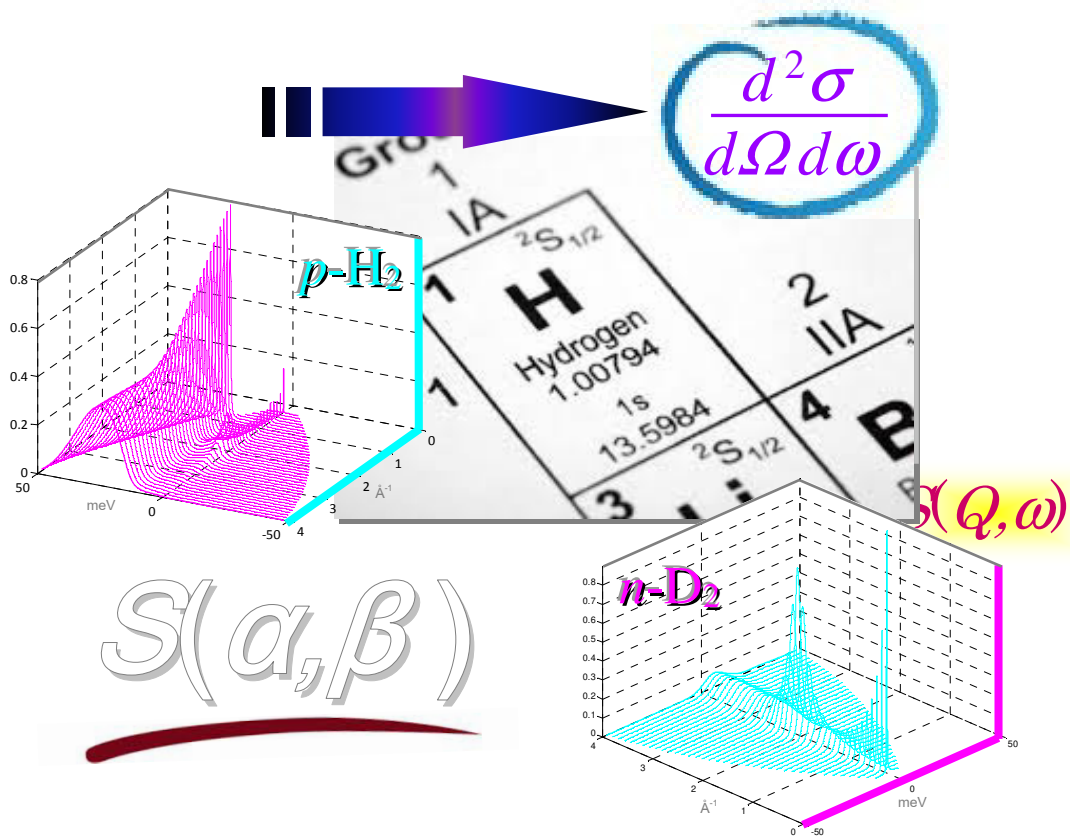

**Institut Laue Langevin**
Grenoble, France

**July 2014**

# CONTENTS

# The neutron cross section of the hydrogen liquids:
## substantial improvements and perspectives

by
E. G<small>UARINI</small>*

## Abstract

The design of moderators and cold sources of neutrons is a key point in research-reactor physics, requiring extensive knowledge of the scattering properties of very important light molecular liquids such as methane, hydrogen and their deuterated counterparts. Inelastic scattering measurements constitute the basic source of such information but are difficult to perform, the more so when high accuracy is required, and additional experimental information is scarce. The need of data covering as large as possible portions of the kinematic *Q-E* plane thus pushes towards the use of computable models, validated by testing them, mainly, against integral quantities (either known from theory or measured) such as spectral moments and total cross section data. A few recent experiments demonstrated that, at least for the self contribution, which dominates in the incoherent scattering case of hydrogen, accurate calculations can be performed by means of quantum simulations of the velocity autocorrelation function. This method is shown here to be by far superior to the use of standard analytical models devised, although rather cleverly, for generic classical samples. The neutron dynamic structure factor (and consequently the well-known $S(\alpha,\beta)$) of parahydrogen and deuterium, suitable for use in packages like NJOY, are given and shown to agree very well with total cross section measurements and expected quantum properties.

*Permanent affiliation:
Dipartimento di Fisica e Astronomia, Università degli Studi di Firenze, Via G. Sansone 1, I-50019 Sesto Fiorentino (FI), Italy  -  e-mail: guarini@fi.infn.it

Temporary affiliation (November 2013-July 2014): Institut Laue Langevin, Grenoble, France

# I. Introduction

Determinations of dynamic neutron cross sections have lately been recognized as indispensable for improved reactor physics calculations and modern moderator design. In particular, accuracy in the treatment of the thermalization process and the detailed knowledge of the dynamic (i.e., momentum and energy dependent) response to thermal neutrons of moderating materials have become fundamental requirements in this field [1]. However, available international neutron cross section libraries [see e.g. 2] suffer from several drawbacks, mainly related to the approximate methods used to predict the wave vector dependence of the scattering law [3,4]. In the most important cases for neutron moderation purposes, i.e. those of low-mass classical and quantum molecular liquids such as water, hydrogen and their heavy twins ($D_2O$ and $D_2$), these and further approximations more severely compromise the results, and, especially for the slowest neutrons, even the accuracy of the most recent libraries (like JEFF3.1 and ENDF/B-VII.0) turns out to be rather poor [5-8].

For these reasons the present WP of the CRISP project focussed on the opportunity to significantly improve the international reference neutron cross-section databases by direct exploitation of dynamic structure factor $S(Q,E)$ determinations, based either on specific neutron measurements or on experimentally-tested molecular dynamics simulations. The dynamics of crucial systems like water and heavy water has first been considered, with encouraging results showing the feasibility and reliability of this approach [9]. This opened the way to explore, in a similar way, the more complex case of the hydrogen liquids, which was tackled only recently and whose results constitute the central body of this report.

In the second section the basic formulas of neutron scattering and the formalism appropriate for quantum homonuclear diatomic liquids as hydrogen and deuterium are shortly recalled. The typical approximations used in the evaluation of the double differential cross section (DDCS) of these systems are discussed, along with their limitations. The links among the neutron DDCS, the dynamic structure factor $S(Q,E)$, and the dimensionless quantity $S(\alpha,\beta)$, needed as input of well-known nuclear data processing codes as NJOY [3], are also directly explicitated.

A survey of the available (extremely limited) inelastic neutron scattering data for $H_2$ and $D_2$ is reported, showing the overall lack of reliable $S(Q,E)$ measurements of immediate utility for the scopes of this project.

In the absence of suitable experimental information, the quantum simulation-based route to *self* DDCS calculations is here (Sect. III) shown to be a very effective tool to obtain the scattering behaviour of liquid *para*-$H_2$ in a vast region of the kinematic plane, as confirmed by the excellent agreement of our computations both with total cross section (TCS) existing data and with the true *quantum* value of the spectral second frequency moment, obtained from *thermal* to *hot* neutrons without *any adjustment* of the dynamical parameters.

The fourth section provides the evidence that, differently from the para-$H_2$ case, the knowledge of the centre-of-mass (CM) *self* dynamic structure factor is in no way sufficient to predict the scattering cross section of liquid $D_2$, except in the kinetic regime of nearly free-gas behaviour (the approximations standardly used to model the additional *distinct* contributions to the total DDCS are discussed in Sect. II). Since the signal due to intermolecular space and time correlations is proportional to the coherent scattering properties of the system, its effect is crucial, in *most* kinematic conditions, for the deuterium case. Nonetheless, its role is also evident in the scattering of *cold* neutrons (< 15 meV, i.e. with energies unable to induce rotational transitions) from p-$H_2$. The difficulties in both the theoretical modeling and simulation of the coherent collective dynamics of these quantum liquids clearly point at the need to resort, when appropriate, to accurate experimental determinations of the scattering law, preferably carried out on time-of-flight instruments.

Finally, the last section reports the presently achieved DDCS results which are of ready and reliable use for the production of nuclear-data files in the *.ACE* (A Compact ENDF) format required by neutron transport codes as MCNP [10] and McStas [11]. Conclusive remarks about the significant improvements introduced by this work and on the further experimental and quantum simulation activities still deserved by the hydrogens in the cold and ultra-cold neutron range, can be found at the end of this report.

## II. The double differential cross section of $H_2$ and $D_2$

Thermal scattering evaluations stored in the various releases of the Evaluated Nuclear Data Files (ENDF) libraries and used in modern reactor-analysis codes are typically generated by means of the NJOY Nuclear Data Processing System [3,4]. Various modules compose the NJOY package, which is meant to produce thermal scattering laws and cross sections for the most common moderating materials at selected temperatures. The last modules in the NJOY chain, which are used to obtain *.ACE* files, require as input the scattering law given in a special form, widely known as $S(\alpha,\beta)$. The latter is actually a *dimensionless equivalent* of the (dimensioned) neutron DDCS, and it is a function of the *dimensionless* variables $\alpha$, $\beta$ related to wave vector and energy transfers $Q$, $E$, respectively. The correspondence between the "languages" used in nuclear physics and in neutron scattering studies of condensed matter will be clarified in the following subsection.

Apart from amending some specific limitations (see next sections) of the available scattering kernels for the hydrogen liquids [see e.g. 12-15], the innovatory idea behind this project is to run NJOY to produce nuclear-physics-compliant files, but using input $S(\alpha,\beta)$'s either derived from direct DDCS neutron measurements, or from refined scattering kernels exploiting modern classical or quantum molecular dynamics simulations of the dynamic structure factor $S(Q,E)$. In technical words, this means that the LEAPR [4] module of NJOY is actually skipped, and replaced by experiments or by differently managed (i.e., outside the NJOY package) calculations; both ways being more reliable than LEAPR in producing input $S(\alpha,\beta)$'s for the THERMR [4] module.

The wide-spread terminology in this field, linking $S(\alpha,\beta)$ to the neutron DDCS and to $S(Q,E)$, strongly and implicitly refers to the concepts of neutron scattering from *monatomic* systems. Differences are thus worth clarifying when dealing with moderating *molecular* liquids. The next subsection aims to provide the necessary definitions and glossary of reference while limiting, at the same time, the frequent ambiguities and unclear use of symbols that can be encountered in the literature. The specific case of hydrogens will be shortly summarized.

### II.1 Basic definitions

In nuclear neutron scattering from liquids the accessible experimental quantity is the double differential cross section (DDCS) per scattering unit:

$$\frac{d^2\sigma}{d\Omega\, dE} = \sqrt{\frac{E_1}{E_0}}\; S_n(Q,E)$$

where $E_1$ and $E_0$ are the scattered and incident neutron energies, respectively, and $S_n(Q,E)$ is the neutron-weighted combination of the *self* and *distinct* components of the dynamic structure factor $S(Q,E) = S_{self}(Q,E) + S_{dist}(Q,E)$. The dynamic structure factor is the space and time Fourier transform of van Hove's density-density space- and time-dependent pair correlation function $G(r,t)$ [16]. When written as a function of the neutron energy transfer $E = E_0 - E_1 = \hbar\omega$, the dynamic structure factor $S(Q,E)$ has units of an inverse energy (e.g. meV$^{-1}$) and, coming from the theory of correlation functions, it is exclusively a system-dependent quantity that nothing has to do with the probe used to measure it. Differently, $S_n(Q,E)$ depends on the neutron probe we are specifically using, despite neutronists often call it "$S(Q,E)$" as well. In particular, for a *monatomic and monoisotopic* system, we have

$$S_n(Q,E) = \frac{\sigma_{coh}}{4\pi} S(Q,E) + \frac{\sigma_{inc}}{4\pi} S_{self}(Q,E) = \frac{\sigma_{coh}}{4\pi} S_{dist}(Q,E) + \frac{\sigma_b}{4\pi} S_{self}(Q,E)$$

where $\sigma_{coh}$ and $\sigma_{inc}$ are the bound coherent and incoherent scattering cross sections of the isotope under consideration, and $\sigma_b = \sigma_{coh} + \sigma_{inc}$ is its total scattering cross section. If $S(Q,E)$ is measured in meV$^{-1}$, then $S_n(Q,E)$ is in barn sr$^{-1}$ meV$^{-1}$, i.e. has the same units of a DDCS and not of the true dynamic structure factor (meV$^{-1}$).

In the case of diatomic *molecular* systems that approximately behave as free vibro-rotors, like the hydrogens down to liquid temperatures, it is possible to write $S_n(Q,E)$ in terms of the translational centre-of-mass (*CM*) dynamics and the roto-vibrational quantum numbers ($J_0, J_1, v_0=0, v_1$), as

$$S_n(Q,E) = u(Q) S_{CM,dist}(Q,E) + \sum_{J_0 J_1 v_1} F(Q, J_0, J_1, v_1) S_{CM,self}(Q, E - E_{J_0 J_1} - E_{v_0 v_1}) \qquad (1)$$

where $u(Q)$ is an appropriate $Q$-dependent function containing the coherent cross sections of the nuclei in the molecule, and the function $F$ takes different expressions according to the nuclear spin statistics and the ortho-para concentration, and contains both the coherent and incoherent scattering cross sections [17 and Refs. therein]. In particular, the second term represents the intramolecular dynamics as a sum of spectral lines centred at the various energies of the roto-vibrational transitions. Note that Eq. (1) defines a *molecular* quantity.

Incoherent scattering dominates the response from liquid H$_2$, thus in the kinematic conditions where intense rotational transitions are excited (like the 0 $\rightarrow$1 one, above 14.7 meV), the neutron $S_n(Q,E)$ actually coincides with the *self* (intramolecular) part of Eq. (1):

$$S_n(Q,E)\big|_{H_2} \xrightarrow{\text{thermal and hot neutrons}} \sum_{J_0 J_1 v_1} F(Q, J_0, J_1, v_1) S_{CM,self}(Q, E - E_{J_0 J_1} - E_{v_0 v_1}) \ .$$

Therefore, having a good model line-shape for the CM *self* dynamics is enough to guarantee the knowledge of the whole DDCS of H$_2$ in a vast kinematic region, without the need of consuming experiments.

In both the mono- and diatomic cases here considered $S_n(Q,E)$ actually represents the (neutron) response function, therefore it is an asymmetric function of $E$ obeying the detailed balance condition. Switching to the $S(\alpha,\beta)$ language, care must be taken about the definitions reported in the literature. Here the formalism of Ref. [4] will be adopted, therefore the following definitions hold for the DDCS in terms of the *asymmetric* $S(\alpha,\beta)$ per atom:

$$\frac{d^2\sigma}{d\Omega\, dE} = \frac{\sigma_b}{2 k_B T}\sqrt{\frac{E_1}{E_0}}\; S(\alpha,\beta)$$

with ($M$ is the mass of the scattering unit)

$$\alpha = \frac{\hbar^2 Q^2}{2 M k_B T} \qquad \beta = -\frac{E}{k_B T} = \frac{E_1 - E_0}{k_B T}$$

Clearly, the $S(\alpha,\beta)$'s are simply an adimensional and monatomic-like equivalent of the neutron $S_n(Q,E)$'s, and therefore of the DDCS. In the case of a homonuclear diatomic molecule we have:

$$S(\alpha,\beta) = \frac{k_B T}{\sigma_b} S_n^{(molec)}(Q,E)$$

where $\sigma_b$ is the bound scattering cross section of one nucleus, while the superscript "*molec*" indicates that $S_n(Q,E)$ corresponds to the molecular expression of Eq. (1).

Finally, it is useful to recall that calculations of the DDCS according to Eq. (1) require a choice for the CM dynamic structure factor $S_{CM}(Q,E) = S_{CM,self}(Q,E) + S_{CM,dist}(Q,E)$. These are discussed in the following subsection.

## II.2 Analytical models for the translational dynamics and their limitations

II.2.1 The *self* CM dynamics

The *self* component in $S_{CM}(Q,E)$, most important for incoherent scatterers as $H_2$, is usually modeled in two ways [4, 12-15]. The first is the ideal gas (IG) law [see e.g. 17] giving rise to the well-known Young and Koppel model [18] for the hydrogen *self* DDCS; the other is the original Egelstaff and Schofield (ES) model [19], though duly modified to comply with detailed-balance asymmetry, and with the first frequency moment sum rule that ensures translational spectra centred at the (non-zero) recoil energy $E_r = \hbar^2 Q^2/2M$, with $M$ the molecular mass. In particular, the so-modified ES model can be written as

$$S_{CM,self}^{(ES)}(Q,E) = \frac{1}{\hbar\pi} D Q^2 \tau \lambda \frac{\exp(DQ^2\tau) K_1\left(\tau\lambda\sqrt{\omega^2 + (DQ^2)^2}\right)}{\sqrt{\omega^2 + (DQ^2)^2}} \exp\left(\frac{\hbar\omega}{2 k_B T}\right) \quad (2)$$

with

$$E = \hbar\omega,\; \tau = \frac{MD}{k_B T},\; \lambda = \sqrt{1 + \left(\frac{t_0}{\tau}\right)^2},\; t_0 = \frac{\hbar}{2 k_B T},$$

and *D* the self diffusion coefficient, whose temperature dependence can be modeled as in Ref. [20]. The last exponential factor in Eq. (2) guarantees detailed-balance asymmetry and $K_1$ is the modified Bessel function of the second kind. However, it can be shown that the *quantum* second-moment sum rule cannot be fulfilled by the above ES model unless by arbitrarily modifying the parameters (usually taken from experiment, like the self diffusion coefficient) entering the model. In particular, for quantum systems the second moment sum rule is given by [16]:

$$M^{(2)}(Q) = \int dE \, E^2 \, S_{CM,self}(Q,E) = \frac{2Q^2}{3M}\langle E_K \rangle + \left(\frac{E_r}{\hbar}\right)^2 \qquad (3)$$

where $<E_K>$ is the mean kinetic energy of the particle, which generally differs from the classical value $3/2 \, k_B T$. Experimental and Path Integral Monte Carlo simulation values of $<E_K>$ for para-hydrogen and deuterium at various liquid temperatures have been provided by Celli et al. [21] and Colognesi et al. [22]. Fig. 1 shows the comparison of the IG and ES results for the second frequency spectral moment, obtained by integration of the line-shapes over a wide-enough energy range. Clearly the ES model misses both the classical and quantum prescriptions, here corresponding to $T = 15.7$ K.

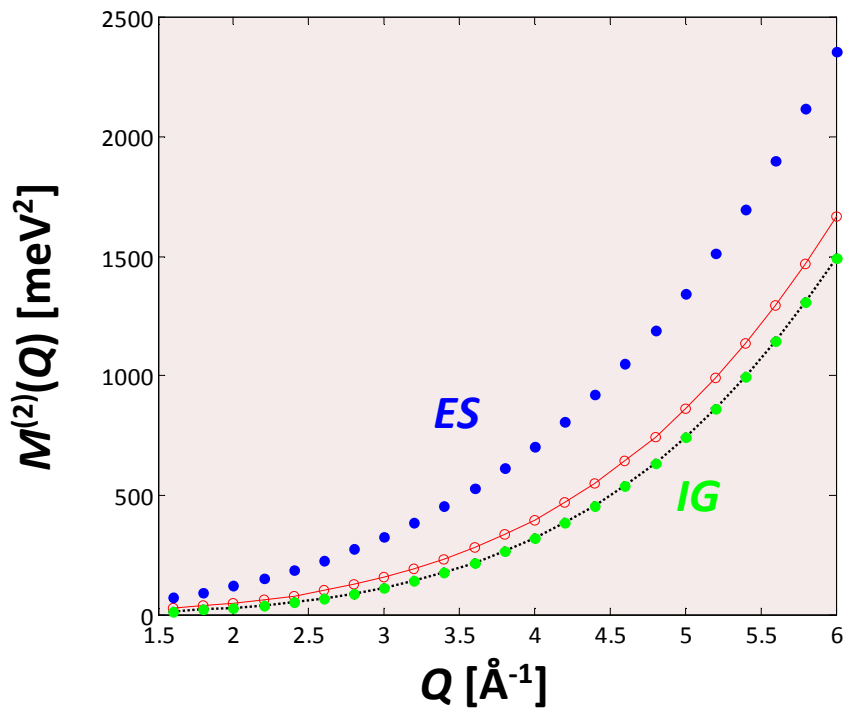

**Fig. 1** - Second frequency moment of p-H$_2$ at 15.7 K. The ES results (blue dots) clearly differ from the expected quantum values (red circles) joined by a continuous curve) as obtained from Eq. (3) using the experimental $<E_k>$ estimate of Ref. [21]. The IG model, while intrinsically respecting the first moment quantum sum rule, follows the $M^{(2)}(Q)$ behaviour (green dots) corresponding to a mean kinetic energy of $3/2 \, k_B T$, typical of a non-interacting system (black dotted line).

It is worth observing that an *ad hoc* increase of $\tau$, for instance through an augmented effective mass, can indeed provide the correct quantum values. However, the required effective mass has no convincing origin and turns out to be much smaller than that used in Ref. [13], and there justified by assuming hindered diffusion in the liquid, with the single molecule actually behaving as a heavier cluster of about 40 molecules. While there is no apparent evidence in favour of the choice of such a large effective mass (except the achievement of a better *fit* of the experimental TCS data), it is anyway sure that, with such mass values, the 'improved' scattering kernel proposed in Ref. [13] is irrespective of at least one of the fundamental requirements imposed by the quantum nature of the hydrogen liquids. Indeed, the neglect of the quantum properties of these low-temperature liquids is common to all existing scattering kernels concerning the hydrogens: all are based on the IG or ES models for the *self* part [4, 12-15] and, apart from adjustments, these are substantially unchanged since two decades, as confirmed by the absolutely identical treatment proposed for $H_2$ and $D_2$ both in Ref. [4] and in the latest release of the NJOY package of 2012 [15].

As shown in Fig. 2, the limitations of both the IG and ES models in representing appropriately the *self* translational dynamics of a true *quantum* dense liquid as $H_2$ are clearly reflected not only in the discrepancies here revealed at the level of $M^{(2)}(Q)$, but more importantly by their inability to reproduce p-$H_2$ TCS data [23] in the crucial range of thermal incident energies ($15 < E_0 < 50$ meV).

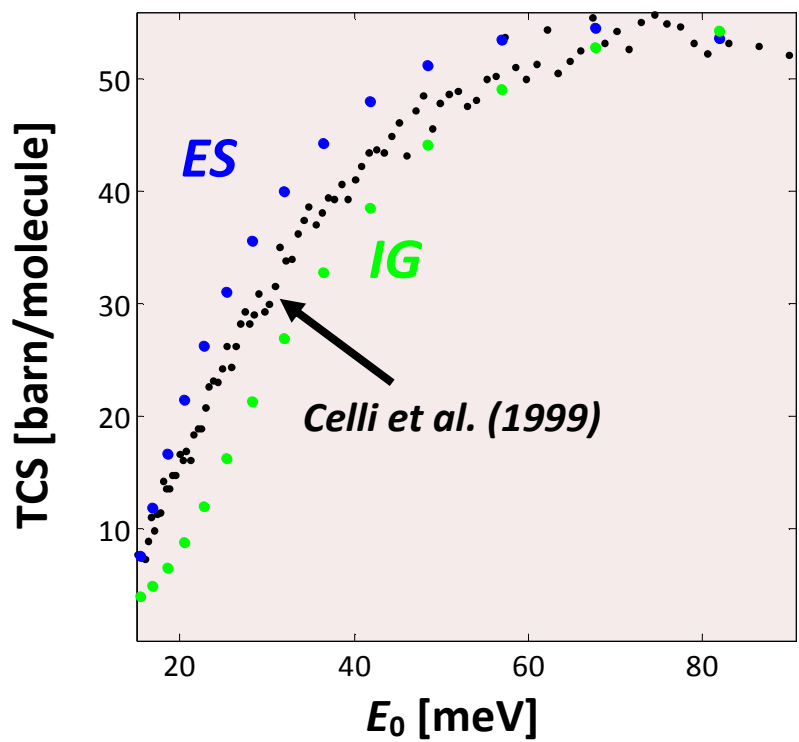

**Fig. 2** - Total scattering cross section of p-$H_2$ at 15.7 K. The IG (green dots) and ES (blue dots) results are compared with the experimental data (black dots) of Celli et al. [23] in the thermal region.

II.2.2 The *distinct* CM dynamics

The schematization of the coherent collective contribution to the DDCS is an even more difficult task, given the general complications arising from the absence of an exact theory for the dynamic structure of liquids as soon as the hydrodynamic regime is abandoned ($Q \rightarrow 0$) and the free-gas behaviour is still far from reached. Indeed, hydrodynamic theory cannot typically be used to model the CM dynamic structure factor at wave vectors above ~0.2 Å$^{-1}$, since most molecular liquids start to show strong deviations from a $Q^2$-dependence of the damping and from a linear $Q$ behaviour of the excitation frequency [see e.g. 24,25]. Thus the modeling of the coherent signal has lately been based on a mean field approximation [12,14] which, however, is unable to account for one of the few inelastic low-$Q$ measurements on liquid p-H$_2$ [26], and scarcely reproduces the n-D$_2$ spectra of Ref. 27, especially at small scattering angles. Such comparisons remain anyway somewhat ambiguous because of the questionable reliability of the quoted neutron data [26,27]: multiple scattering corrections are simply mentioned, but raw sample+container (and unnormalized) spectra are uniquely published, making very difficult any attempt to properly reproduce such experimental data by means of model calculations of the single scattering from the sample alone.

In the absence of new and more detailed experimental information about the collective dynamics of liquid H$_2$ and D$_2$, two analytical models have historically been used [8] which roughly account for the overall features of the TCS of liquid deuterium and improve the description of the total cross section of hydrogen in the low-energy range. The simplest is based on the Vineyard approximation, which models the *total* CM dynamic structure factor $S_{CM}(Q,E)$ as the *self* spectrum modulated by the static structure factor $S_{CM}(Q)$, i.e.:

$$S_{CM,dist}(Q,E) = [S_{CM}(Q) - 1]\, S_{CM,self}(Q,E);$$

the other is the Sköld model [28]:

$$S_{CM,dist}(Q,E) = S_{CM}(Q)\, S_{CM,self}\left(\frac{Q}{\sqrt{S_{CM}(Q)}}, E\right) - S_{CM,self}(Q,E)$$

which has the merit of recovering the correct value of the second frequency moment for classical systems, but becomes of questionable, if not arbitrary, applicability in the quantum case. Certainly, a good knowledge of the static structure would improve the effectiveness of these approximate representations of the intermolecular dynamics, originally conceived for classical monatomic fluids reasonably described by hard-spheres or Lennard-Jones interaction potentials. However, quantum effects considerably deplete the structural features [29] and cannot be easily foreseen unless by means of accurate quantum simulations. Once again, consideration of this problem is completely disregarded in the present NJOY treatment of the hydrogens case, which is still based on obsolete structural information without an experimental basis [15], despite the availability nowadays of neutron diffraction determinations both for n-D$_2$ [30] and for p-H$_2$ [31]. Figures 3 and 4 clearly show the strong deviations between the structural inputs in NJOY (by Keinert and Sax) and the experimental results. Here and in Ref. 13 the modeling of the coherent dynamics has been improved, at least, by including the existing experimental information about the static structure.

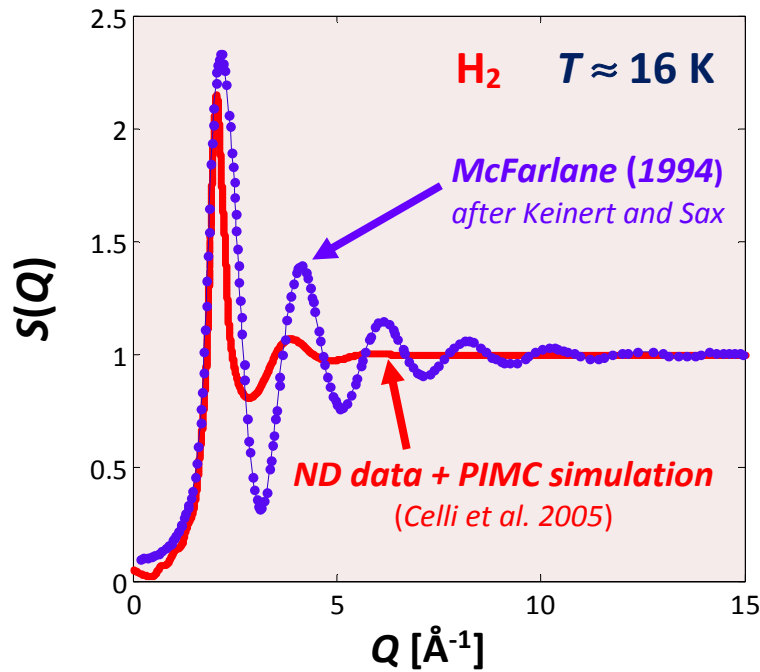

**Fig. 3** - Static structure factor of p-$H_2$ at 16 K as obtained from the merging of neutron diffraction measurements and Path Integral Monte Carlo simulation [31] (red dots). The completely different structure used as input of NJOY calculations [4,15] is shown by the violet curve.

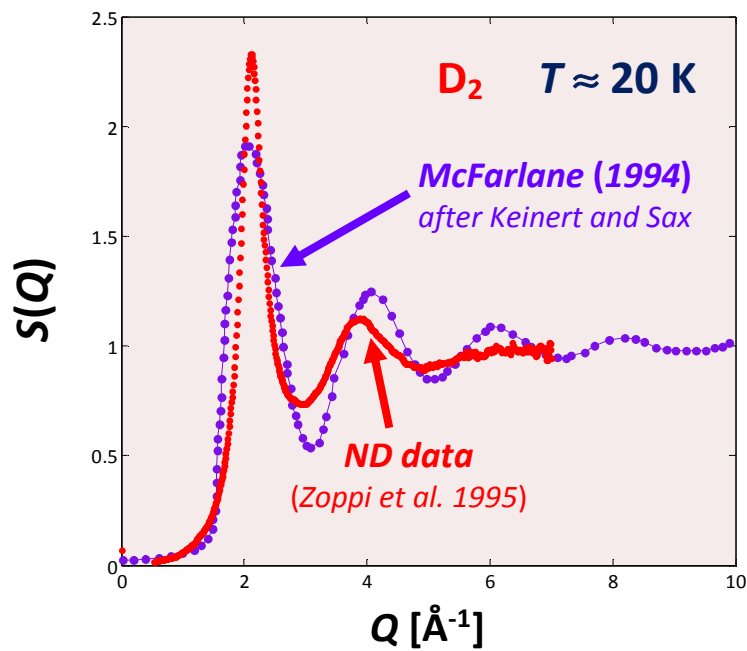

**Fig. 4** - Static structure factor of $D_2$ at 20 K as obtained from neutron diffraction measurements [30] (red dots). The completely different structure used as input of NJOY calculations [4,15] is shown by the violet curve.

## II.3 The experimental situation

As mentioned, inelastic neutron scattering data on the hydrogens are extremely few, despite the general importance of these liquids. Moreover, the accuracy and usability of some of the published DDCS data [26,27,32] seem questionable, due to missing or undetailed corrections of the spectra, like multiple scattering or even container scattering. Serious difficulties arise about the unspecified treatment of the former in years when the importance of the energy-distribution of multiple events in inelastic scattering measurements on liquids was considerably underrated with respect to present times [33].

Figs. 5 and 6 schematically summarize the scarce availability of DDCS neutron data over the kinematic plane, for $H_2$ and $D_2$ respectively. Available $Q$ values are displayed as a function of the maximum energy transfer characterizing each data set, i.e. as a function of the incident neutron energy $E_0$ employed in the measurements.

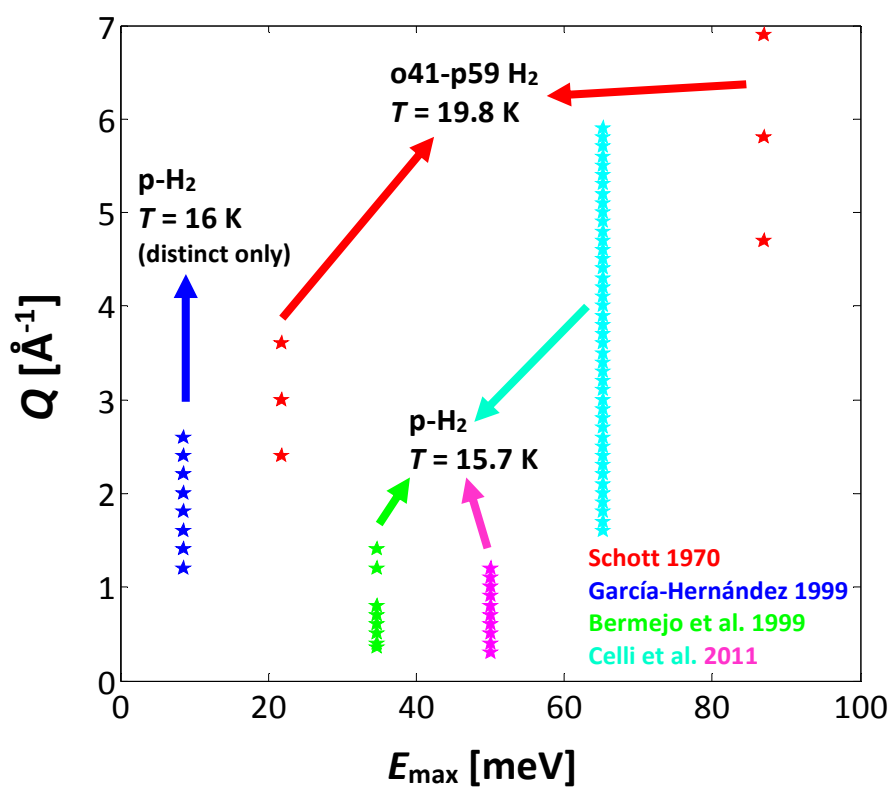

**Fig. 5** - Wave vector transfer values covered by inelastic scattering measurements on $H_2$, at different temperatures and ortho-para concentrations, as a function of the experimental incident energy $E_0 = E_{max}$. Most DDCS published data are not sufficiently accurate for the present project. Conversely, the recent data of Celli et al. [36] provided a crucial test of the validity of $S_{CM,self}(Q,E)$ determinations based on quantum CMD simulations of the velocity autocorrelation function [36] and the use of the Gaussian approximation [38,39].

Among those on hydrogen, the recent measurements of the CM self dynamics of p-$H_2$ by Celli et al. [36] were found to be in globally good agreement with calculations performed in the Gaussian approximation (GA) [38,39] using recent Centroid Molecular Dynamics (CMD) simulations for the velocity autocorrelation function. Only small differences revealed a non-Gaussian behaviour of the data in a very limited *Q* range, which can be assumed to be irrelevant for cross-section calculation purposes. The experimental results of Ref. [36] thus provided a convincing proof of the efficiency of CMD quantum simulations used in combination with the GA for the prediction of the hydrogen self dynamics. Such a method (see Sect. III) represents therefore an extremely valid alternative to most experiments or to the use of "quantum-insensitive" analytical models for $S_{CM,self}(Q,E)$ in the DDCS formula of $H_2$ and $D_2$.

A great advantage of this approach is that simulations of the velocity autocorrelation function only depend on the temperature, and can thus be used for isothermal calculations of $S_{CM,self}(Q,E)$ wherever wished in the kinematic plane. This actually means that, at least for $H_2$, *we are presently able to calculate the DDCS, and consequent $S(\alpha,\beta)$, in most kinematic conditions and with unprecedented accuracy*.

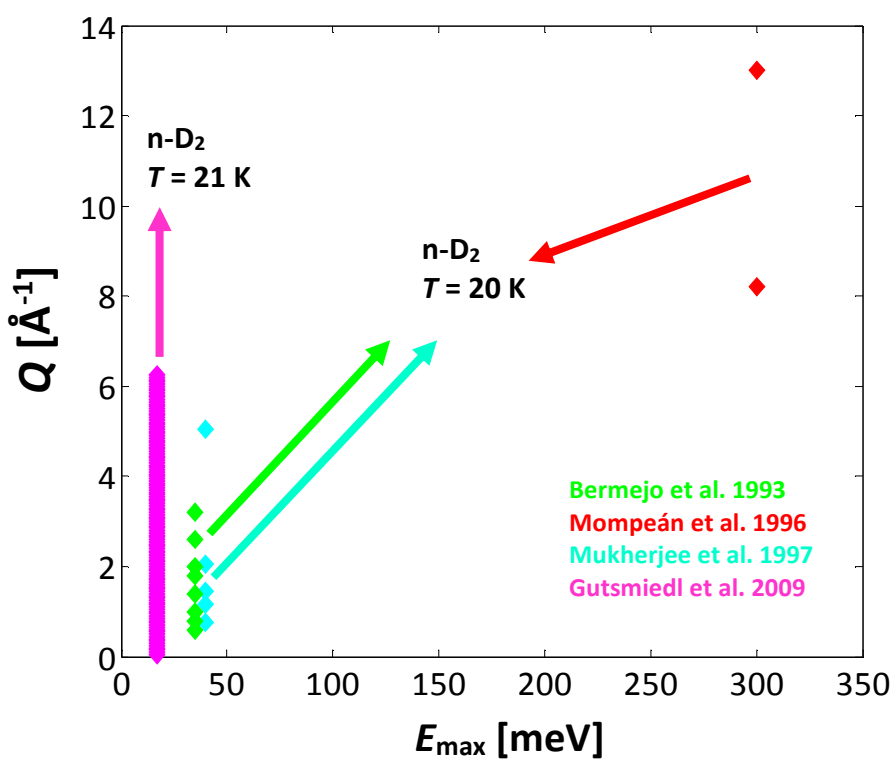

**Fig. 6** - As in Fig. 5, but for liquid $D_2$.

As shown in Fig. 6, experiments probe an even smaller portion of the thermal neutron range in the $D_2$ case. As for $H_2$, a reliable use of the deuterium spectra published in the nineties is forbidden by the lack of details on data analysis [27,40], while the liquid-phase results of a more recent experiment on solid $D_2$ [41], though available on the ILL database, require a full analysis exceeding the terms of this project.

# III. Improved description of hydrogen *self* dynamics

As stated previously, the IG and ES models of the self dynamic structure factor, used until present in DDCS computations [3,4,12-15], do not accurately describe the quantum dynamics of liquid hydrogen and, consequently, its response to neutrons $S_n(Q,E)$. A clearly superior account of the quantum properties of the spectral lineshape, and, via double (over energy and solid angle) integration of Eq. (1), of the measured TCS for para- and normal $H_2$ [23,42] is instead given by the quantum-simulation-based approach discussed in the previous section, here adopted and proposed for the first time as a valid alternative to the typically skill-demanding and time-consuming experiments on the hydrogen liquids. In the following, the results of this method will synthetically be denoted as *q-simul+GA*.

## III.1 Quantum Centroid Molecular Dynamics (CMD) at work

Despite their importance and partial successes, path integral (PI) CMD and similar methods (e.g. ring polymer molecular dynamics) do not capture the full quantum character of a many-body molecular system, as far as exchange effects and treatment of quantized rotations are concerned. Fortunately, in the case of liquid para-$H_2$, these deficiencies have no consequences, since quantum exchange was shown to be irrelevant [43] and quantum rotations are included via the Young and Koppel [17,18] formalism, which only depends on $S_{CM,self}(Q,E)$. Indeed, almost all published quantum simulation results on liquid parahydrogen, based on such assumptions, provide results in satisfactory agreement with the experimental measurements of mean kinetic energy, diffusion coefficient, and structural properties.

The PI CMD method was applied to a system of 256 molecules interacting via the Silvera-Goldman potential [44]. The Trotter number, i.e., the number of beads on the classical ring polymers replacing the quantum mechanical particles, was 64. In contrast to the usual implementation, the calculation of the quantum mechanical forces, which are required at each time step of the otherwise classical simulation, was performed by the path integral Monte Carlo method, rather than MD, thus avoiding sampling problems associated with the stiff "intramolecular" modes of the polymers and allowing for a much larger time step. The simulation was extended up to 1 ns in the isokinetic ensemble, ensuring thermal stability and statistical reliability. The velocity correlation was calculated up to a maximum time lag of 1.5 ps. A shorter test run with 500 particles confirmed that the shape of the velocity autocorrelation function (VACF) was not noticeably influenced by finite-size effects.

The dynamical information conveyed by the VACF is a keypoint in the development of models for the *self* part of the DDCS of viscous dense fluids. In particular, the Gaussian approximation provides $S_{self}(Q,E)$, once the frequency spectrum of the VACF is known. For hydrogen, the width function of the ES model (related to the VACF) has been typically adjusted [12,13,20] to obtain a reasonable agreement with available cross-section experimental results. For example, the VACF obtained in Ref. [12] for $H_2$ at 14.7 K is reported in Fig. 7. Quite expectedly, such a "classical" and indirect determination differs significantly from the output of the present quantum simulations, also shown in the figure.

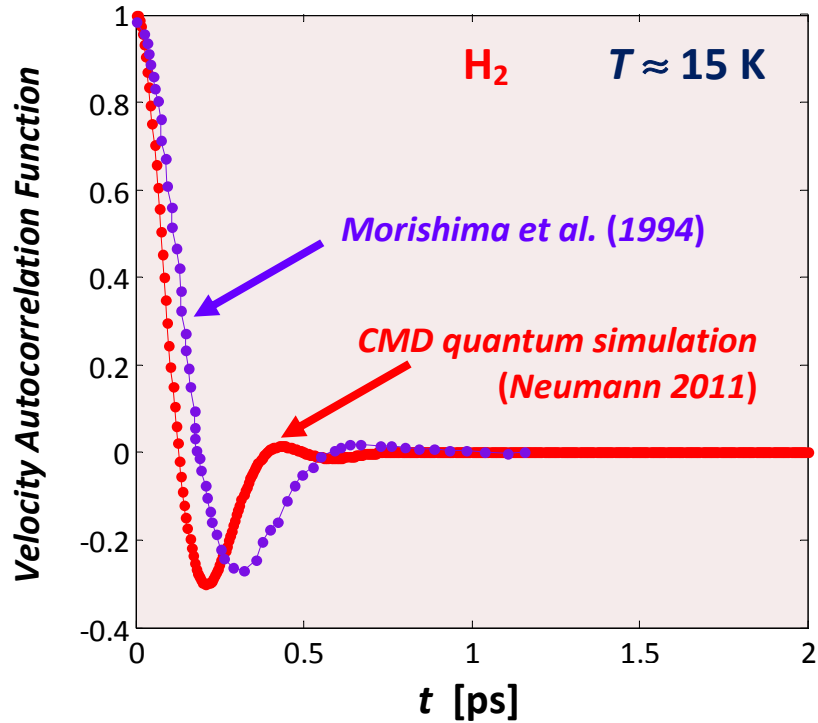

**Fig. 7** - Velocity autocorrelation function of hydrogen deduced in Ref. [12] (violet dots). The quite different VACF resulting from quantum CMD simulations [36] is shown in red.

III.1.1 Results in the Gaussian approximation

The output of a PI CMD simulation is the *canonical* (or Kubo-transformed [45]) VACF:

$$u_c(t) = \frac{1}{\beta} \int_0^\beta d\lambda \left\langle e^{\lambda H} \mathbf{v}_{CM}(0) \, e^{-\lambda H} \mathbf{v}_{CM}(t) \right\rangle$$

where $H$ is the hamiltonian operator of the system and $\beta = 1/k_B T$. The *self* intermediate scattering function $F_{CM,self}(Q,t)$ in the Gaussian approximation can then be written as [39]:

$$F_{CM,self}(Q,t) = \exp\left\{ -\frac{E_r}{\hbar} \int_0^{+\infty} d\omega \, \frac{f(\omega)}{\omega} \left[ [1-\cos(\omega t)] \coth\left(\frac{\beta \hbar \omega}{2}\right) - i\sin(\omega t) \right] \right\} \quad (4)$$

with

$$f(\omega) = \frac{M\beta}{3\pi} \int_{-\infty}^{+\infty} dt \, e^{-i\omega t} \, u_c(t) \, .$$

The self dynamic structure factor is then obtained as the time Fourier transform of Eq. (4) at each desired $Q$. The $Q$-dependence of the second frequency moment in the *q-simul+GA* case, compared in Fig. 8 with the IG and ES results already shown in Fig. 1, is in far better agreement with the theoretical prescription of Eq. (3).

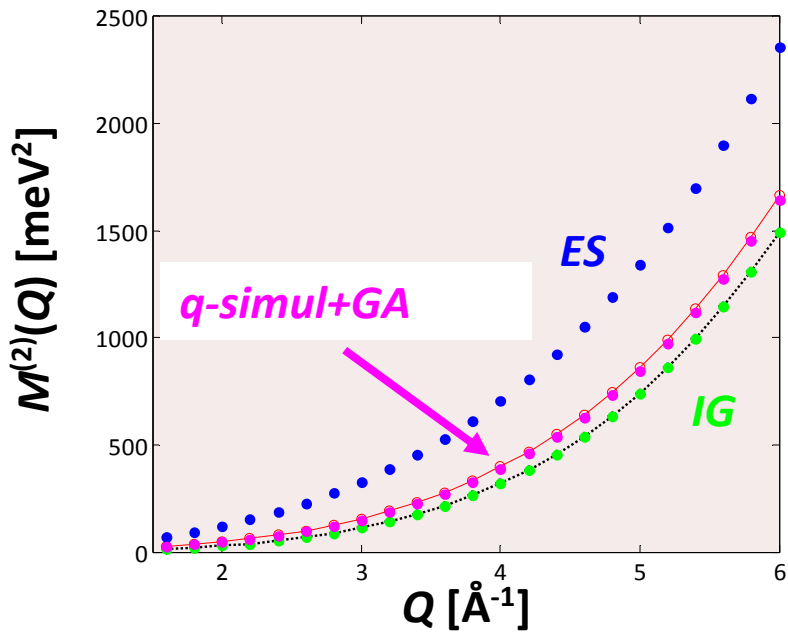

**Fig. 8** - As in Fig. 1 but including the *q-simul+GA* results (pink dots).

The superiority of the *q-simul+GA* calculations is evident also at the level of the p-$H_2$ TCS data reported in Fig. 9. Agreement is excellent wherever the *self* component dominates the dynamics of $H_2$ ($E_0 > 10$ meV), even before ideal-gas behaviour is reached by any model.

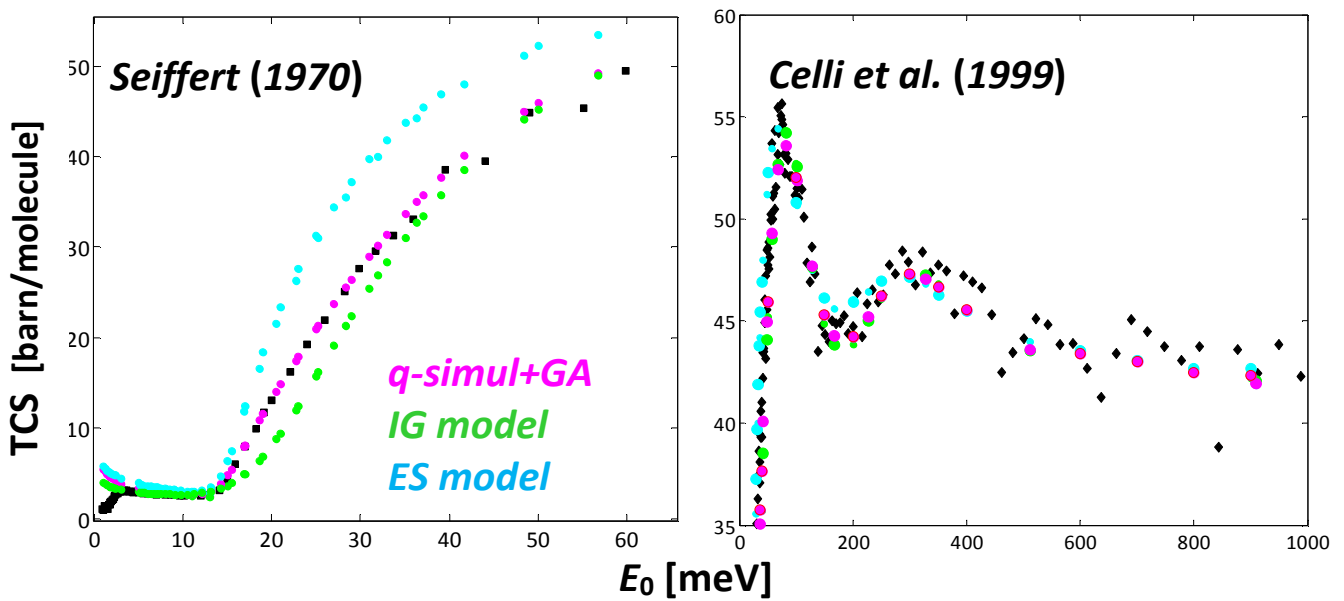

**Fig. 9** - Total cross section per molecule of p-$H_2$ at 15.7 K in the cold and thermal range covered by Seiffert measurements [42] and at the higher energies reached in the spallation-source experiment of Celli et al. [23].

An equally good agreement is obtained for the TCS of normal $H_2$, as displayed in Fig. 10.

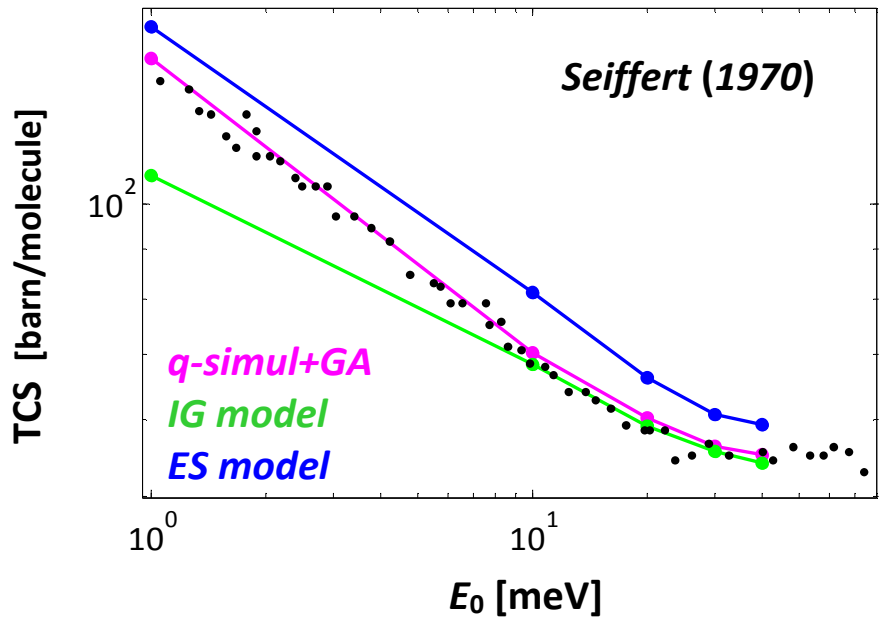

**Fig. 10** - Total cross section per molecule of n-$H_2$ at 15.7 K in the cold and thermal range covered by Seiffert measurements [42]. Note the logarithmic scale.

At the more detailed level of non-integrated quantities, the effectiveness of the q-simul+GA method can be tested against the DDCS data obtained by Schott for $H_2$ at 19.8 K in the 41-59 ortho-para concentration [34]. In this case the CMD simulations available at 21.2 K were employed. Our *q-simul+GA* results (Fig. 11) are superior to those of Ref. [14] and uncomparably better than those of Schott himself (using the IG model) and of Utsuro [46].

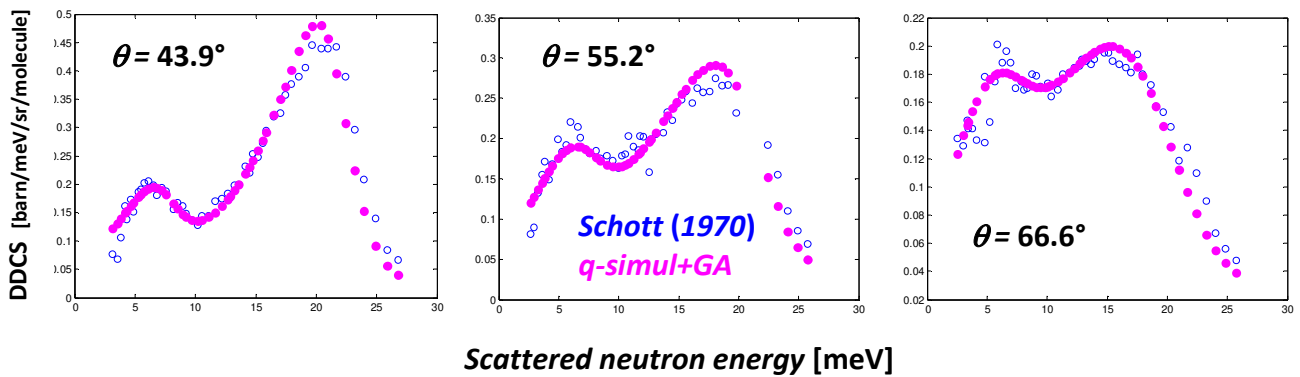

**Fig. 11** - Experimental DDCS of 41o-59p $H_2$ at 19.8 K (blue circles) [34] and *q-simul+GA* data (pink dots). Measurements refer to the thermal range ($E_0$ = 21.8 meV) and rather wide angles indicated in each frame, where hydrogen dynamics is fully dominated by the *self* component.

# IV. Coherent scattering effects

The results of the preceding section demonstrate that q-simul+GA-based calculations can substitute experiments on $H_2$ at thermal incident energies (20-100 meV) and medium-high Q values (20-onwards Å$^{-1}$). Actually, there is no need to 'measure', but simply to 'calculate' the *self* DDCS, spanning the needed kinematic region with extreme flexibility and reliability. Nonetheless, the few (both DDCS and TCS) data available on para-$H_2$ indicate the importance of considering the collective contributions too: both at small incident energies (below 10 meV), where sound modes are not excited in liquid hydrogen ($c_s$ > 1200 m/s at 15 K) but thermal diffusion mirrors the many-particle nature of the system not yet overwhelmed by the intramolecular dynamics, and at slightly higher incident energies, where both sound and thermal-diffusion in- and quasi-elastic collective modes add up to growing rotational self-molecule contributions.

Collective phenomena, probed by the coherent neutron signal, obviously are enhanced in the response of liquid $D_2$. Thus, even a good and quantum-compliant representation of the *self* dynamics, as the q-simul+GA one, cannot anyway be sufficient to account for the overall neutron scattering properties of deuterium which, differently from the $H_2$ case, does not benefit from the simplifications induced (at least at certain incident energies) by a huge incoherent-to-coherent ratio.

Unfortunately, quantum simulation methods have not yet been shown to provide a convincing estimate of the total (*self* plus *distinct*) dynamic structure factor. Indeed, CMD is unreliable as soon as the reference operators are not linear in the particle positions and momenta, like those appearing in the intermediate scattering function $F(Q,t)$ [47]. Tentative simulations of the collective behaviour of a quantum coherent liquid as $D_2$ can in fact be observed to be in quite a poor agreement with experiment [48], although tests of the simulations certainly suffer from the lack of reference inelastic measurements and from the mentioned uncertainties on the few existing ones [27,37]. In the persisting absence of both reliable simulations and measurements of the total $S_{CM}(Q,E)$, the present possibilities to foresee the full $H_2/D_2$ response to neutrons can only pass through the use of the approximate recipes discussed in Sect. II.2.2 and the test against TCS experimental data.

### IV.1 Low-energy response of liquid p-$H_2$

At cold neutron energies the coherent component plays a role in p-$H_2$ too. This is evident in Fig. 9, where any calculation of the *self* part progressively departs from the (absorption-corrected) TCS measurements of Seiffert [42] as soon as the incident neutron energy is decreased below 10 meV. Such discrepancies are considerably reduced by quitting the "self approximation" and recovering the full expression of $S_n(Q,E)$ given in Eq. (1). The *distinct* contribution to scattering was thus taken into account by means of either the Vineyard or the Sköld approximation, using in both cases the experimental $S_{CM}(Q)$ of Ref. [31]. Figure 12 shows the effect of the addition of either of the two modelings of the distinct part to the self one obtained in the q-simul+GA case.

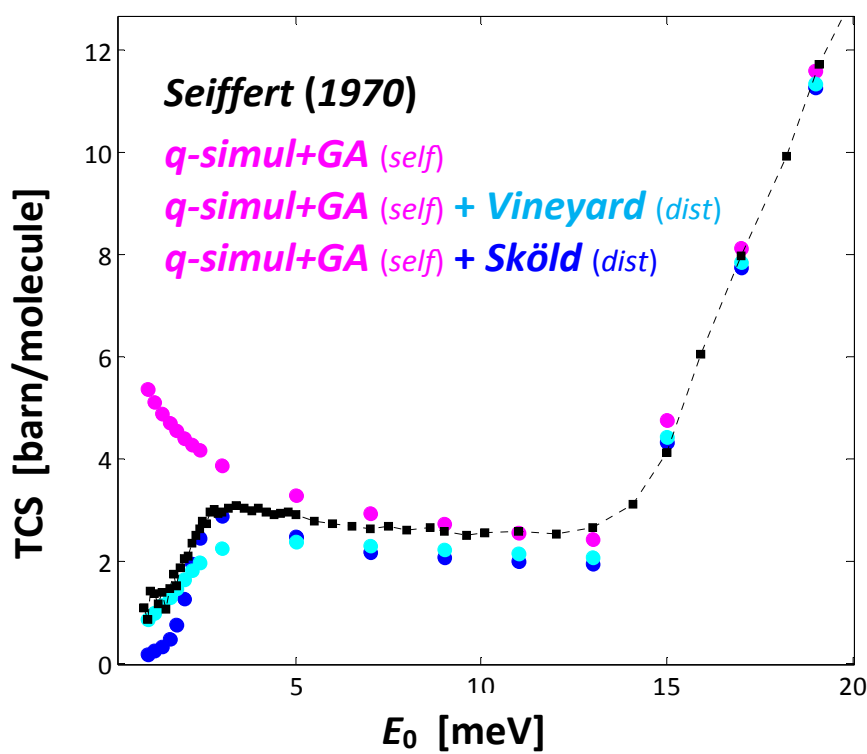

**Fig. 12** - Total scattering cross section of p-$H_2$ at 15.7 K at cold and thermal energies. The *q-simul+GA* values (pink dots) are compared with the absorption-corrected experimental data (black squares) of Seiffert [42]. The other displayed results correspond to the addition of a distinct contribution in the Vineyard (cyan dots) and Sköld (blue dots) approximation.

Quite surprisingly, discrepancies are visibly reduced only below ~ 2 meV, where the Vineyard approximation better combines with the q-simul+GA *self* result than the Sköld one. Actually, such approximate models for the distinct contribution are of arbitrary applicability in the quantum case, and certainly do not guarantee any full compliance with the quantum second-moment sum rule. Nonetheless, a very good account of the measured TCS between 1 and 12 meV is instead obtained by combining the ES model for the self part with the Vineyard version of the *distinct* dynamics, as shown in Fig. 13. This remains anyway an unexplained result, likely due to a fortuitous combination of the inefficiencies of both the self (ES) and distinct (Vineyard) modelings, which apparently (and effectively) cancel out in an integrated quantity as the total scattering cross section.

    The previous results confirm the need of accurate measurements of the DDCS of liquid p-$H_2$ able to sensitively probe the different contributions to the total signal in the low-energy range below ~ 15-20 meV.

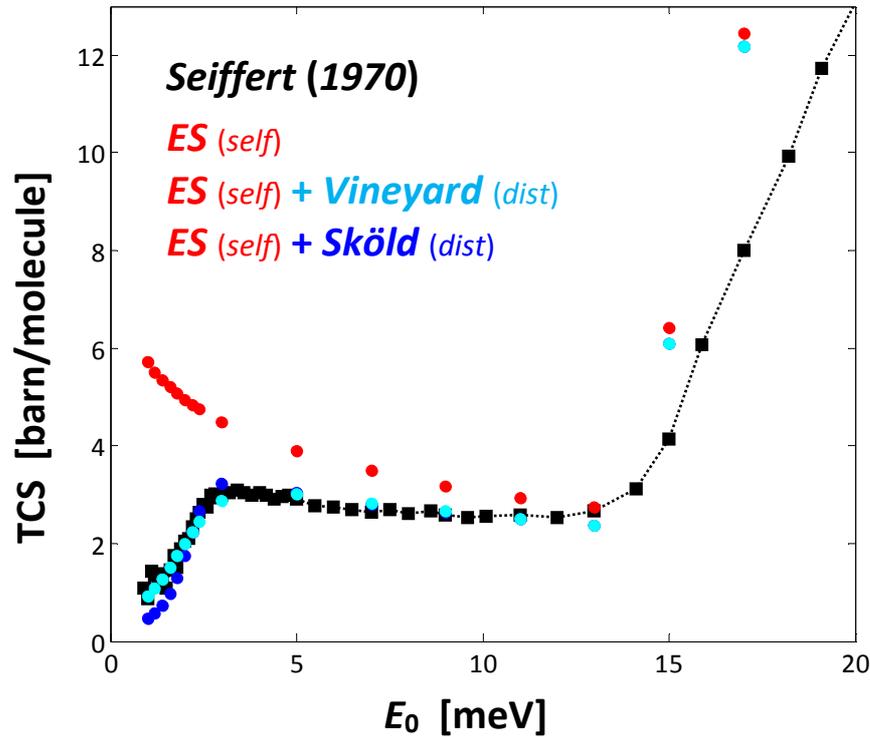

**Fig. 13** - Total scattering cross section of p-H$_2$ at 15.7 K at cold and thermal energies. The *ES* values (red dots) are compared with the absorption-corrected experimental data (black squares) of Seiffert [42]. The other displayed results correspond to the addition of a distinct contribution in the Vineyard (cyan dots) and Sköld (blue dots) approximation.

### IV.2 The more demanding case of liquid D$_2$

Differently from hydrogen, coherent scattering affects the total cross section of liquid deuterium at all incident energies accessed by experiment, presently reaching 80 meV at maximum [42]. Neglect of the distinct component in Eq. (1) leads to a considerable overestimate of the TCS measured values. Figure 14 shows how the ES or IG self models clearly miss an appropriate description of the data in the whole experimental range, while a more realistic energy dependence of the TCS is obtained by including the intermolecular dynamics in the Vineyard or Sköld forms and exploiting the experimental information on the static structure [30].

In the deuterium case the Vineyard approximation is globally less effective than the Sköld one, the latter recovering the correct TCS values between 1 and 2 meV independently of the used self model (IG or ES). However, a deeper inspection reveals that the IG + Sköld model is unable to account for the further details of the TCS data, like the shape and height of the peak centred at about 3.5 meV. A better agreement is obtained with the ES-based calculations reported in Fig. 15, where the effects of the Vineyard and Sköld models are also compared. Beyond the overall agreement with the data, the ES + Sköld combination is anyway unsatisfactory in the peak region from 2.5 to 4 meV. In summary, none of the analytical models for $S_{CM}(Q,E)$ describes the details of the TCS even in the case of D$_2$: once again, an attempt through the quantum simulation route thus appeared to be mandatory.

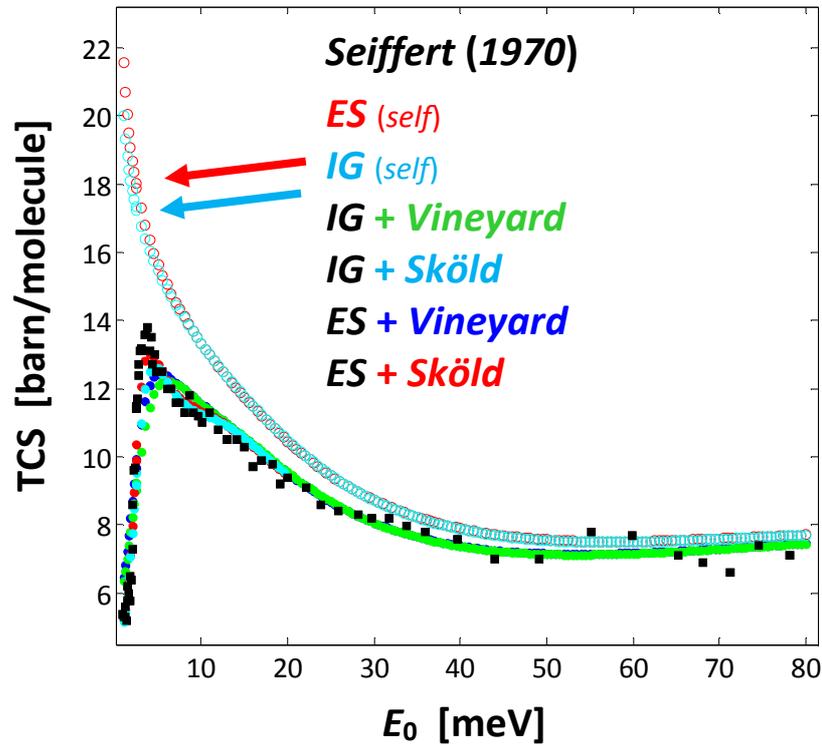

**Fig. 14** - Total scattering cross section of n-$D_2$ at 19 K at cold and thermal energies. The *ES* and *IG* values (red and cyan) circles are compared with the experimental data (black squares) of Seiffert [42]. The other displayed results correspond to the addition of a distinct contribution in the Vineyard (green and blue dots) and Sköld (cyan and red dots) approximation.

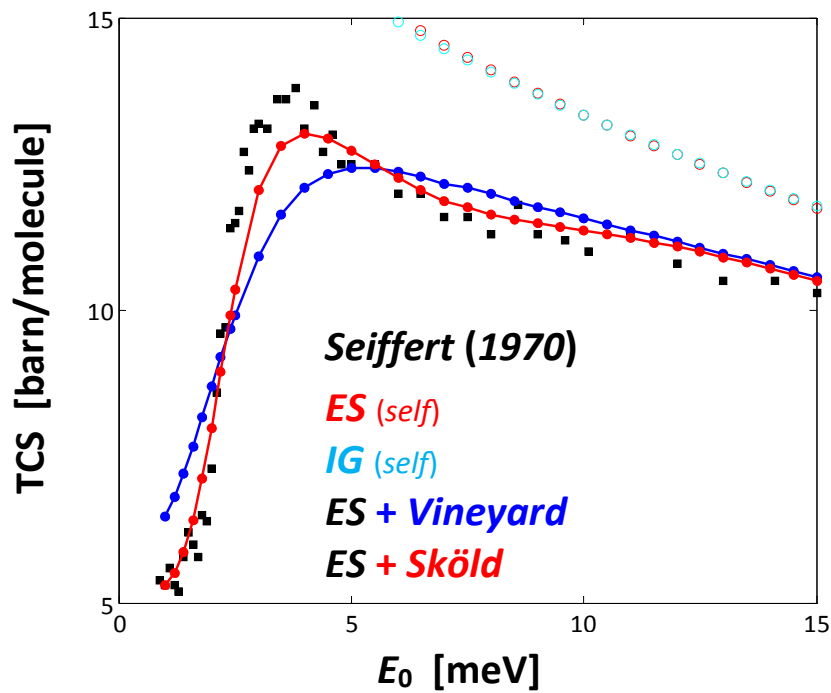

**Fig. 15** - Total scattering cross section of n-$D_2$ at 19 K at cold and thermal energies. The *ES* + Vineyard (blue) and *ES* + Sköld (red) results are compared with the experimental data (black squares) of Seiffert [42].

Preliminary CMD simulations for liquid deuterium at 20 K have been available to us only very recently, in the final phases of this project. Although calculations of the corresponding total cross section are quite lengthy, we are able to present here the first results concerning the most demanding energy range, *i.e.* that lying below 10 meV.

As for hydrogen, the self dynamic structure factor of $D_2$ at 20 K was calculated in the Gaussian approximation (Sect. III.1.1) starting from the Kubo transform of the simulated VACF. The q-simul+GA *self* line-shape was then used in the calculation of the DDCS according to Eq. (1), following the Vineyard or Sköld approximations to evaluate the *distinct* contribution. The results are shown in Figs. 16 and 17, for the Vineyard and Sköld modelings respectively. In each figure we also report the IG- and ES-based calculations for comparison.

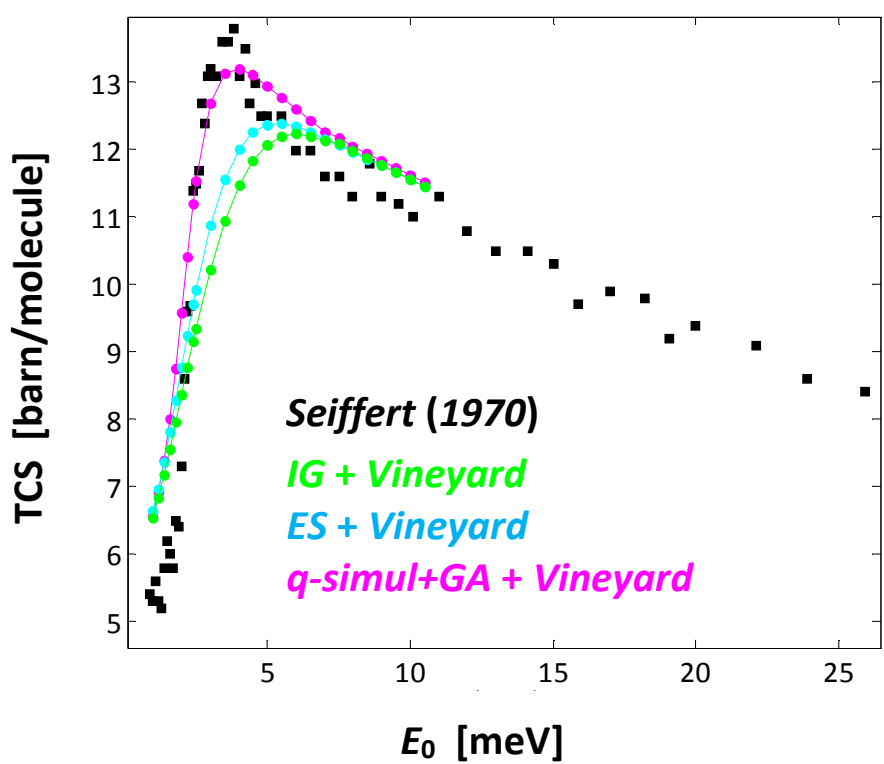

**Fig. 16** - Total scattering cross section of n-$D_2$ at 19 K (black squares) [42]. The Vineyard schematization of the *distinct* dynamics is combined with the IG (green dots), *ES* (cyan dots) and very recent q-simul+GA (pink dots) results for the *self* component.

A strong reduction of the deviations from experiment is achieved whatever distinct model is used in combination with the quantum simulation results. In particular, a very good description of the data is obtained in the q-simul+GA + Sköld case reported in Fig. 17. Residual deviations from the measured cross section apparently correspond to a limited constant offset which is difficult to ascribe to inaccuracies of the simulations.

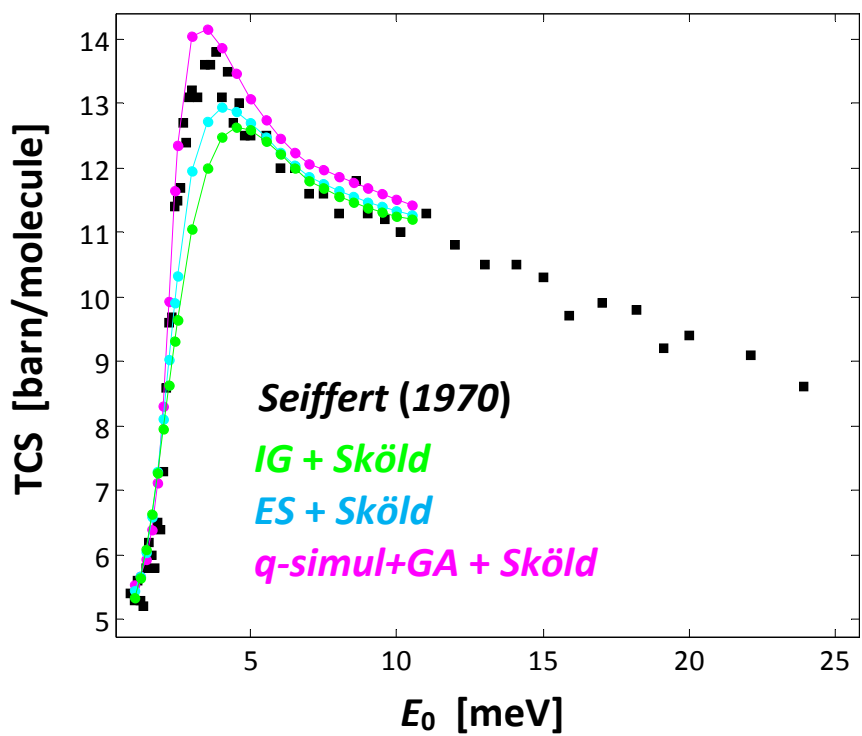

**Fig. 17** - Total scattering cross section of n-$D_2$ at 19 K (black squares) [42]. The Sköld schematization of the *distinct* dynamics is combined with the IG (green dots), *ES* (cyan dots) and very recent q-simul+GA (pink dots) results for the *self* component.

The present findings indeed confirm the efficiency of the simulation-based DDCS calculations for both hydrogen liquids. The limited quantum behavior of liquid deuterium might explain the partial success of the Sköld model, provided a quantum-compliant *self* line-shape is taken anyway as the starting point of any calculation.

### IV.3 Final results

Confident of the validity of the q-simul+GA representation of the hydrogens CM self dynamics, we performed DDCS calculations for para-$H_2$ and normal $D_2$ in various conditions (variable incident energy and $Q,E$ range). Figs. 18 and 19 give an example of the results that can be easily obtained, with extremely limited CPU time (few minutes on a standard personal computer), by means of our Matlab code implementing Eq. (1). In particular, we report some example spectra obtained at 35 meV incident energy, including the distinct term in the Vineyard and Sköld form for $H_2$ and $D_2$, respectively.

The corresponding contour plots in the kinematic region spanned by the previous example calculations are reported in Fig. 20 and 21. In Fig. 20, the quasi-elastic peak, as well as the dominant 0->1 rotational line of p-$H_2$, can be easily discerned.

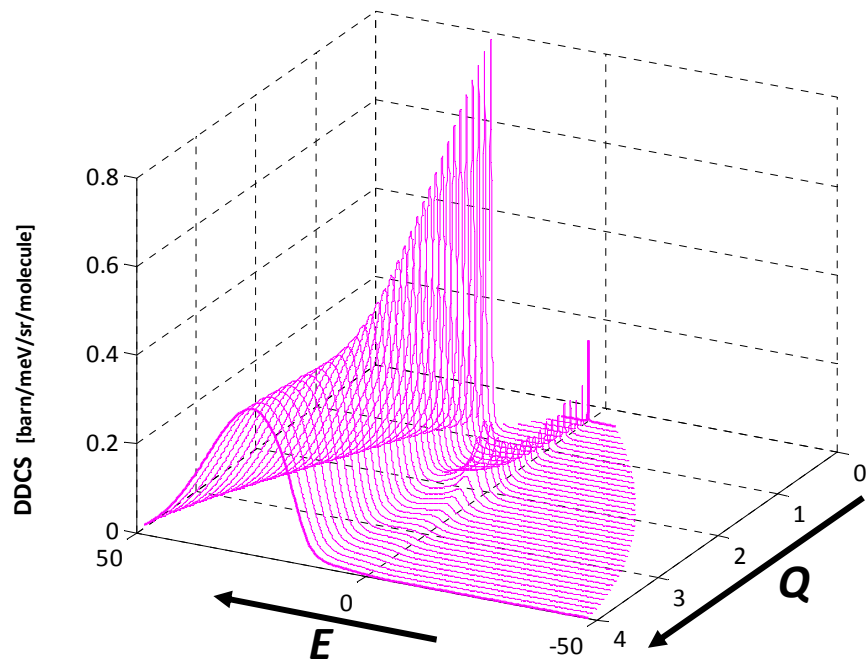

**Fig. 18** - Double differential cross section of p-H$_2$ at 15.7 K and 35 meV incident energy. The Vineyard schematization of the *distinct* part was combined with the q-simul+GA result for the *self* term. The rotational 0→1 line dominates the spectra at most *Q* values.

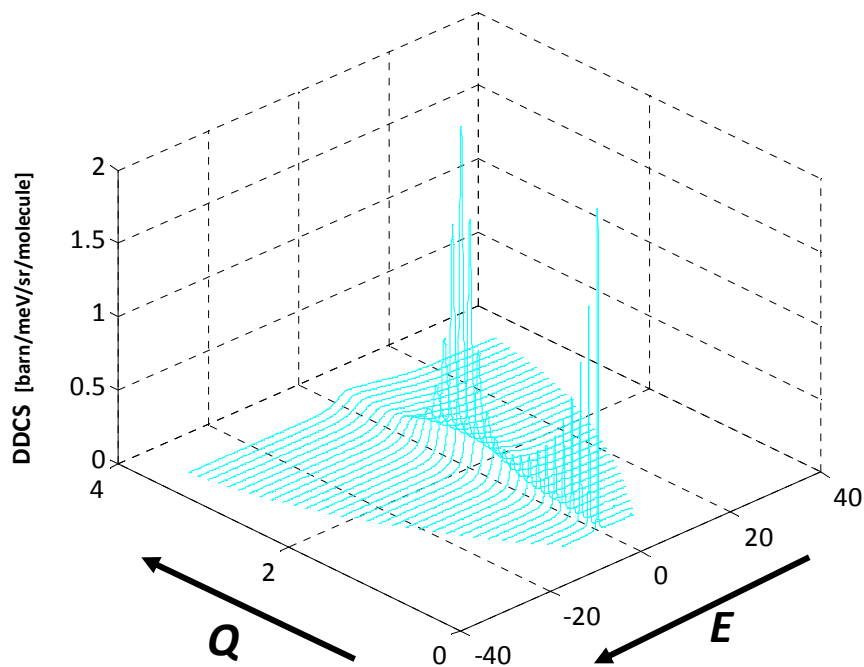

**Fig. 19** - Double differential cross section of n-D$_2$ at 20 K and 35 meV incident energy. The Sköld schematization of the *distinct* part was combined with the q-simul+GA result for the *self* term.

Following the definitions of Sect. II.1, conversion of DDCS data (per atom) to $S(\alpha,\beta)$ input files appropriate for nuclear data processing codes as NJOY is straightforward.

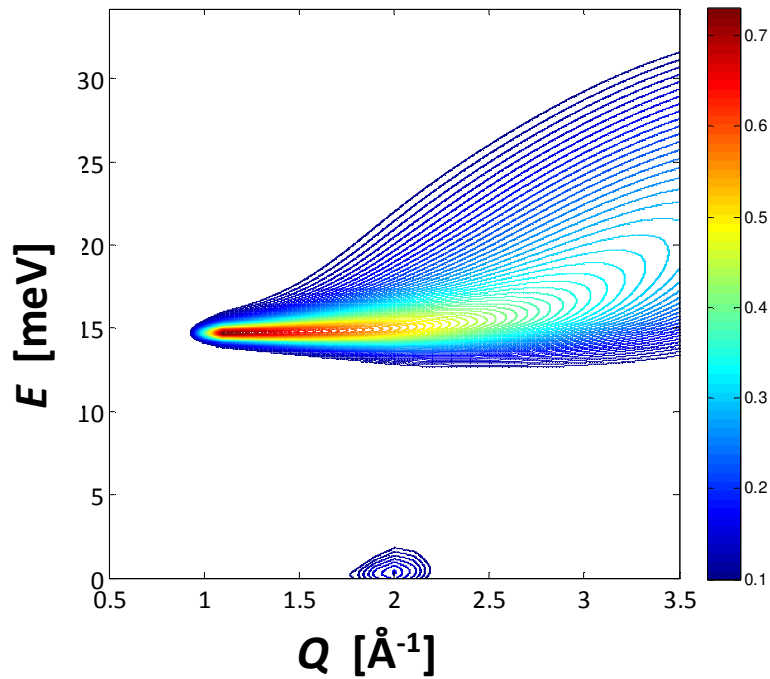

**Fig. 20** - Contour plot corresponding to the double differential cross section data of p-$H_2$ at 15.7 K and 35 meV incident energy displayed in Fig. 18.

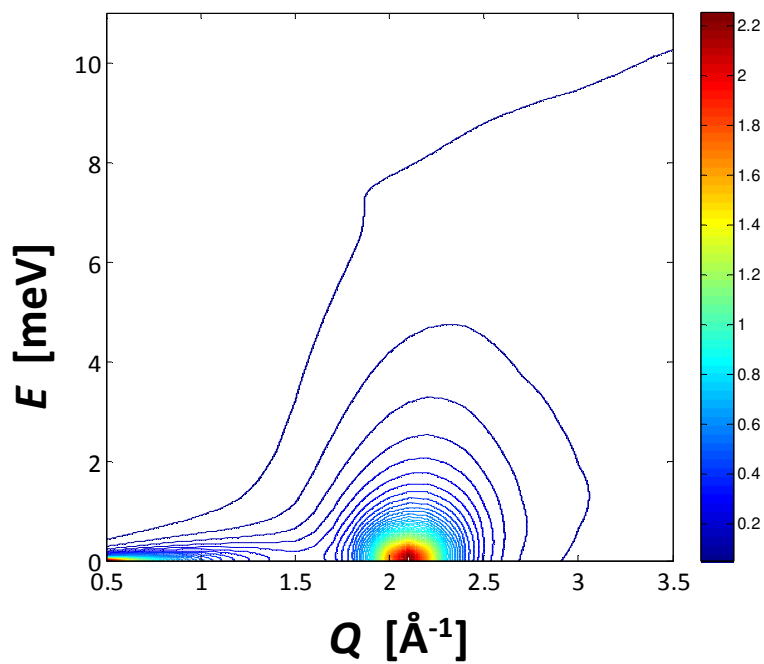

**Fig. 21** - Contour plot corresponding to the double differential cross section data of n-$D_2$ at 20 K and 35 meV incident energy displayed in Fig. 19.

# V. Conclusions

This work provided quite encouraging results as concerns the possibility to obtain accurate evaluations of the scattering of cold, thermal and hot neutrons from cryogenic liquids as important as $H_2$ and $D_2$. We developed, verified and implemented a powerful method able to accurately account for the quantum behaviour of these fluids and which has the undoubtful merit of limiting considerably the expensive experimental efforts and lengthy data treatment typically required by neutron scattering on the hydrogens. Great part of the kinematic plane can be covered by this technique, though experiments remain of crucial importance for the evaluation, refinement and validation of possible descriptions of the coherent dynamics, and therefore, mainly, for the deuterium case and in the low-energy range.

Although an improved evaluation of the true coherent collective component is envisaged on an experimental basis, the present estimates represent, to our knowledge, the only ones providing an unprecedented agreement with total cross section measurements without resorting to any adjustment of one or more dynamical parameters. Indirectly, this is one of the strongest validations of the CMD technique for the prediction of the VACF of quantum liquids.

Cold neutron moderator design would certainly profit from investments regarding the acquisition of quantum simulation facilities able to easily cover different thermodynamic conditions, thus paralleling, from the thermodynamic point of view, the already fast computation tools set up in the present project for the calculation of the molecular double differential cross section at selected temperatures.

# ACKNOWLEDGMENTS

This work took enormous advantage from the world-recognized experience about the hydrogen liquids of the Italian research group working at the Istituto dei Sistemi Complessi of the Consiglio Nazionale delle Ricerche (CNR) in Florence: I'm indebted to M. Zoppi, D. Colognesi, M. Celli and U. Bafile for extremely helpful advices, discussions and support. The important stepforwards here achieved in the determination of the DDCS of liquid hydrogen and deuterium have been made possible thanks to the unvaluable collaboration of Prof. M. Neumann in Vienna, who provided the CMD quantum simulations of the velocity autocorrelation function of liquid hydrogen and, appositely for this project, of deuterium. I'm extremely grateful to the project responsible Y. Calzavara and collaborators as E. Farhi for their enthusiastic support and involvment, constantly accompanying my work. The warm welcome, professional assistance and always prompt availability of the ILL staff during my whole stay is also gratefully acknowledged.

**CRISP - WP11**
*Moderator neutron cross-section data*
The neutron cross section of the hydrogen liquids:
*substantial improvements and perspectives*31